\documentclass[twocolumn,trackchanges]{aastex701}

\begin{document}

\title{Simulated LSST Observations of Real Metre-scale Imminent Impactors}

\author[orcid=0009-0009-1340-7081,sname='Frazer']{Michael A. Frazer}
\affiliation{Space Science and Technology Centre, Curtin University, Kent St, Bentley 6102, Boorloo/Perth, WA, Australia}
\affiliation{International Centre for Radio Astronomy Research, Curtin University, Kent St, Bentley 6102, Boorloo/Perth, WA, Australia}
\email[show]{michaelfrazer117@gmail.com}  

\author[orcid=0000-0001-9226-1870,sname='Devillepoix']{Hadrien A. R. Devillepoix}
\affiliation{Space Science and Technology Centre, Curtin University, Kent St, Bentley 6102, Boorloo/Perth, WA, Australia}
\affiliation{International Centre for Radio Astronomy Research, Curtin University, Kent St, Bentley 6102, Boorloo/Perth, WA, Australia}
\email{hadrien.devillepoix@curtin.edu.au}  

\author[orcid=0000-0003-1955-628X,sname='Deam']{Sophie E. Deam}
\affiliation{Space Science and Technology Centre, Curtin University, Kent St, Bentley 6102, Boorloo/Perth, WA, Australia}
\affiliation{International Centre for Radio Astronomy Research, Curtin University, Kent St, Bentley 6102, Boorloo/Perth, WA, Australia}
\email{sophdeam@gmail.com}  

\begin{abstract}   
As of mid-2026, 11 objects have been discovered prior to impacting the Earth, with warning times between 2 - 20 hours. Using real metre-sized Earth impactors from the last decade, we ask the question: ``If the Vera C. Rubin Observatory's Legacy Survey of Space and Time (LSST) had been operating over the last decade, how many imminent impactors would it have observed and discovered pre-impact, and how early would these discoveries have been?" We use the LSST Solar System Survey Simulator \texttt{Sorcha} and a population of real fireballs observed by orbital sensors over the last decade to investigate which events would have been observed pre-impact. We find that the LSST would have observed 30 (13.9\%) of the 216 simulated objects, with most objects receiving 2 - 4 observations. Using the default linking algorithm, only two (0.9\%) of these objects would have been `discovered' pre-impact. Using a modified linking algorithm better suited to fast moving objects, this increases to eight (3.7\%). Based on this, we predict that the LSST will discover 8 $\pm$ 2 imminent impactors over its nominal 10 year survey, at the low end of previous estimations. However, we predict these objects to be discovered $\sim$4 days pre-impact, substantially earlier than the current average. This will bring significant opportunities for telescopic follow-up, targeted fireball observations, planetary defence planning, and public engagement. There is also significant potential for precovery for impactors observed by the LSST but discovered by other surveys, instantly lengthening observation arcs and thereby reducing the orbital and impact location uncertainties. In some cases, these observations may also enable the linkage of telescopic observations with observed fireballs post-impact, providing valuable pre-impact astrometric and photometric data. This has significant implications for both asteroid research and planetary defence.
\end{abstract}

\keywords{\uat{Asteroids}{72} --- \uat{Fireballs}{538} --- \uat{Near-Earth Object}{1092}}


\section{Introduction}\label{sec:intro}
Each year, the Earth is impacted by 35 - 40 metre-scale objects \citep{Brown02}. While most ablate away entirely in the upper atmosphere, some make it through the atmosphere and land as meteorites, allowing us to directly sample the Near Earth Asteroid (NEA) population \citep{Borovicka15}. Others, with sufficient mass and certain entry conditions - such as the 2013 Cheylabinsk event \citep{Borovicka13, Brown13} - can create airbursts with shock waves capable of causing significant damage to people and infrastructure \citep{Tapia17}.

Both situations - research and planetary defence - benefit from observing these metre-scale objects prior to impact. For research, constraining the pre-impact orbit, size, density, and colour/composition of the object and comparing this to any recovered meteorites creates a valuable link between the telescopically-observed asteroid and the lab-analysed hand sample \citep{Jenniskens25, Egal25}. For defence, warning of an imminent impact and the ability to to constrain its location and energy can inform the civil response, particularly through specific planetary defence organisations such as the International Asteroid Warning Network (IAWN; \citealt{Koschny24, Brown13, Devillepoix19, Wheeler24, Bolin25}). 

In the case of large (D $>$ 140 m) Near Earth Objects (NEOs) like (99942) Apophis, this potential warning time can be on the order of years \citep{Giorgini08}, while for smaller (1 - 10 m) NEOs, warning times are typically $<$ 24 hours \citep{Egal25}.

As of February 2026, there have been 11 asteroids/meteoroids that have been discovered in space prior to impacting Earth's atmosphere\footnote{https://cneos.jpl.nasa.gov/pi/} (e.g. \citealt{Jenniskens09, Egal25}), and another which was subsequently found in historical telescope data once the impact had occurred \citep{Clark23}. These metre-scale objects have been detected 2 - 20 hours before impact, typically at apparent magnitudes of $\sim$ 18 - 20 upon discovery, and down to $\sim$ 13 closer to the impact\footnote{https://www.minorplanetcenter.net/mpec/K23/K23CA3.html}.

Some events have been subject to substantial follow-up observation campaigns by other observatories, allowing a unique opportunity to link asteroidal photometry and spectroscopy with meteorite recovery (e.g. 2008 TC$_3$; \citealt{Jenniskens09}) and/or fireball imagery (e.g. 2022 WJ$_1$, 2024 BX$_1$; \citealt{Kareta24, Spurny24}).

These (as well as other close approaches) have been primarily discovered by non-dedicated (e.g. \citealt{Sarneczky22, Norton26}) and dedicated NEO surveys including the Catalina Sky Survey (CSS; \citealt{Drake09}), the Asteroid Terrestrial-impact Last Alert System (ATLAS; \citealt{Tonry18}), and the Panoramic Survey Telescope and Rapid Response System (Pan-STARRS; \citealt{Chambers16})\footnote{NEO discovery by survey is reported by the Center for Near-Earth Object Studies (CNEOS) (https://cneos.jpl.nasa.gov/stats/site\_all.html)}. Another NEO-hunting telescope - Flyeye - is currently under development by the European Space Agency (ESA; \citealt{Cibin19, Fohring24}).

The recently completed Vera C. Rubin Observatory is a next-generation facility which hosts the 8.4 m Simyoni Survey Telescope and 3.2 gigapixel LSSTCam camera. Rubin's Legacy Survey of Space (LSST) is planned to begin in 2026, building a 10 year time-lapse of the southern sky \citep{Ivezic19}. The LSST is expected to be a powerhouse for Solar System discovery, increasing the number of known main belt asteroids, Jupiter Trojans and trans-Neptunian objects by factors of 5-10 per class \citep{Juric23, Kurlander25}. It will also bring significant change in the NEO population, discovering thousands of previously-unknown NEOs and achieving just over 70\% completeness of potentially hazardous asteroids (PHAs) with D $>$ 140 m and minimum orbital intersection distances $<$ 0.05 AU \citep{Jones18, Kurlander25}. It will also discover hundreds of metre-scale NEOs \citep{Kurlander25}. 

This increase in the number of known of NEOs should also result in an increase in the number of meteoroids observed prior to impacting the Earth. For example, recent work by \citet{Cheng26} finds that LSST would discover $\sim$80\% of large (D $>$ 140 m) impactors and $\sim$10\% of 10 - 20 m impactors, of a synthetic impactor population.

Here, we ask: based on real fireball data, how many metre-scale impactors do we predict the LSST will observe and discover over its 10 year survey?

In Section \ref{sec:data}, we describe the real fireball dataset, collected by U.S. Government sensors (USG; \citealt{Tagliaferri94}), and \texttt{Sorcha}, the LSST Solar System survey simulator (\citealt{Holman25, Merritt25}). We perform one `best case' and one `realistic' survey, based on the colours of the input population. Section \ref{sec:results} details our results from the simulated `best case' survey, describing which of the impactors are observed and linked by the current LSST pipelines. In Section \ref{sec:discussion}, we discuss the more general outcomes of the survey and briefly consider the implications for meteoroid science and civil defence warning times. In Section \ref{sec:predictions}, we discuss a more `realistic' survey (based on input colours) and use that to make predictions for what the LSST will discover over the course of its 10 year operations.

\section{Data}\label{sec:data}
\subsection{USG}\label{sec:USG}
The fireball data is available from the Centre for Near Earth Object Studies (CNEOS) online database\footnote{https://cneos.jpl.nasa.gov/fireballs/}. The data includes records of the location, altitude, total impact energy, and velocity of high-energy fireballs and bolides in the atmosphere. This data is collected by space based U.S. Government (USG) sensors designed to monitor the Earth and its atmosphere for nuclear tests \citep{Tagliaferri94}. USG provides near-global coverage \citep{Brown02}, allowing higher-energy events to be observed more often than ground based camera networks, which only observe a few percent of the planet's surface (\citealt{Devillepoix20, Jenniskens25}). The USG system is sensitive to events with impact energies $>$ 0.05 kT TNT\footnote{1 kT TNT = 4.185 $\times$ 10$^{12}$ J}, which generally corresponds to objects with D $\geq$ 1 m at typical Earth-impact speeds of 11.2 - 70 km s$^{-1}$ \citep{Brown02, Devillepoix19}. 

Our LSST simulation runs from 2015 January 01 - 2024 December 31, so we select USG-observed events from the same window. There are 372 USG events in this 10 year period, 220 of which have velocity data. We discuss how this affects our results more in Section \ref{subsec:total_predictions}. We determine the pre-impact orbits for these events based on \citet{JansenSturgeon19}, and remove 4 objects with negative semi-major axes (implying incorrect orbit determination), leaving 216 events.

\subsubsection{USG Data Issues}\label{subsec:usg_issues}
Previous comparisons of USG data with observations from ground-based fireball observatories of the same event have revealed some limitations within the USG dataset, particularly in the recorded velocity vector (which affects the derived orbits; \citealt{Devillepoix19, Brown23, PenaAsensio24, PenaAsensio25, Chow25}). For some cases, errors have been documented to be as high as 10 km s$^{-1}$ in speed, and up to 90$^{\circ}$ in direction (radiant) \citep{Devillepoix19, Brown23}. 

However, these issues appear to have been resolved from 2018 onwards, with USG observations aligning much more closely with ground-based observations \citep{Brown23, PenaAsensio25, Chow25}. As a mostly statistical study, and considering the majority of our data is from 2018 onwards, we do not expect that this will significantly affect our results.

\subsubsection{Impactor Physical Characteristics}\label{subsec:impactors}
For each event, we use the reported velocity and total impact energy to estimate the mass of the impactor using the kinetic energy relation $E = \frac{1}{2}mv^2$. We assume spherical shape and a bulk density of 1500 kg m$^{-3}$, which we use to calculate a diameter (D) for the object. This maintains consistency with other work (e.g. \citealt{Chow25, Ingebretsen25}), and is based on estimated densities for metre-scale NEAs \citep{Mommert14a, Mommert14b}, small (D $<$ 1 km) NEAs from targeted surveys (e.g. Bennu and Itokawa; \citealt{Hickson18}), and a wider survey from Gaia-derived densities of S-type asteroids \citep{Dziadura23}. 

The distribution of S-type vs C-type NEOs is not well constrained for metre-scale objects, mostly due to NEOs being difficult to observe with spectral surveys. De-biased surveys of larger (D $>$ 100 m) objects suggest an even split between C- and S-type NEOs \citep{Binzel19, Marsset22}, however it has been proposed that thermal stress on metre-scale NEOs may preferentially fracture and remove weaker C-types from the Near-Earth space, contributing to the relative lack of carbonaceous meteorites \citep{Shober25}. Based on this, we perform separate simulations on two populations.

Firstly, we present a `best case' scenario and assume all objects are (relatively) bright S-types. We use this to investigate individual objects, and draw links between physical and orbital parameters and likelihood of discovery. Secondly, we present a more realistic option and assume an even split between C- and S-types, and make our predictions based on that. In the second case, we perform two \texttt{Sorcha} runs - one assuming all C-types, and one assuming all S-types. We then iterate multiple surveys by randomly selecting half S- and half C-types, and make our predictions based on the average of those survey runs.

Albedos ($p_v$) are taken from \citet{Marsset22}, and we use the colours derived for Rubin's bandpasses (and for use with \texttt{Sorcha} specifically) in \citet{Kurlander25} (Table \ref{tab:colours}). Values are presented in Table \ref{tab:colours}. We use the IAU standard HG phase curve model \citep{Bowell89} and set G = 0.15 for all objects.


\begin{table}[htbp]
\centering
\caption{LSST-specific colours and albedos for S- and C-type asteroids as used in this work. From \citet{Kurlander25}.}
\label{tab:colours}
\begin{tabular}{cccccccc}
\hline
Type & $p_v$ & $u-r$ & $g-r$ & $r-i$ & $r-z$ & $r-y$ & $V-r$ \\
\hline
S & $0.24$ & $1.62$ & $0.78$ & $0.50$ & $0.71$ & $0.92$ & $0.35$ \\
C & $0.04$ & $1.28$ & $0.62$ & $0.43$ & $0.72$ & $0.93$ & $0.27$ \\

\hline
\end{tabular}
\end{table}


We can estimate the absolute V-band magnitude $H_V$ for each object using Eq. \ref{eq:diam_to_mag} \citep{Harris97}:

\begin{equation}
\label{eq:diam_to_mag}
    H_V = 5 \times [3.1236 - 0.5 \log_{10}(p_v) - \log_{10}(D)]
\end{equation}

\subsection{\texttt{Sorcha}}\label{Sorcha}
\texttt{Sorcha} is a survey simulator developed for predicting Solar System observations and discoveries with the LSST \citep{Holman25, Merritt25}. \texttt{Sorcha} operates in three main steps - ephemeris generation, observations, and linking.

\subsubsection{\texttt{Sorcha} Ephemeris Generation}\label{subsec:ephemeris}
First, \texttt{Sorcha} ingests the orbital elements of the objects into REBOUND, a N-body particle integrator \citep{Rein12}. REBOUND's extension ASSIST \citep{Holman23} populates a simulation with Solar System objects using the SPICE resources files described in \citet{Acton96} and \citet{Acton18}, and uses its 15th order Gauss-Radau integrator IAS15 \citep{Rein15} to generate a 10-year ephemeris for each test object. 

While \texttt{Sorcha} is a highly accurate software, it (and the underlying REBOUND/ASSIST framework) need to be treated carefully when dealing with close approaches and impactors. Within our version of \texttt{Sorcha}, we set the accuracy parameter to 10$^{-6}$ days, initial timestep to 10$^{-6}$ days, and minimum timestep to 10$^{-9}$ days, which balance accuracy and compute time for close approaches (Chow, I., personal communication). In addition to this, we also increase the texttt{ar\_n\_sub\_intervals} parameter in the configuration file to 500, as suggested for modelling close approaches and Earth impactors\footnote{https://sorcha.readthedocs.io/en/latest/advanced.html}.

See \citet{Holman25} for a more detailed description of \texttt{Sorcha}'s ephemeris generation. 

\subsubsection{\texttt{Sorcha} Observations Generation}\label{subsec:observations}
Once the ephemeris has been generated, the positions of each object are checked against a pointing baseline for the survey. This is composed of the location (pointing), 5-$\sigma$ image depth and observing conditions for the 2.1 billion observations that LSST will make over 10 years, accounting for factors including survey strategy, weather, and scheduled and unscheduled downtime. We use a custom version of the recent v5.0.1 baseline\footnote{https://survey-strategy.lsst.io/baseline/index.html} \citep{Yoachim25}, modified to run from 2015 January 01 - 2024 December 31, allowing us to test against historic fireball observations. This survey strategy uses 30 s exposures in the $grizy$ bands, and 38 seconds in the $u$ band \citep{Yoachim25}. The LSST's general strategy is to observe each location (pointing) twice per night, about 30 minutes apart, in different filters. This generates single-visit limiting magnitudes of around 23.4, 24.6, 24.3, 23.6, 22.9 and 21.7 for the LSST's $ugrizy$ filters \citep{Ivezic19}.

The on-sky position of the object is also compared to a camera footprint file, which can be set either as a circular approximation of Rubin's field of field of view, or the exact shape (including gaps between chips) of the detector. We use the precise LSSTCam detector footprint. If the object is brighter than the 5-$\sigma$ depth for the image and sits within the camera footprint, it is recorded as an observation. 

This process outputs a database of all observations made over the 10 year period. This contains information about the object (on-sky position, apparent trailed source magnitude and associated errors) as well as metadata for the image from the pointing database (including time of observation, on-sky centre of the image, filter, and 5-$\sigma$ depth of the image).

\subsubsection{\texttt{Sorcha} Linking Process}\label{subsec:SSP}
The LSST's Solar System Processing (SSP) pipeline is then applied, which attempts to link individual observations of objects \citep{Myers13, Juric17}. The current SSP links multiple observations of an object on the same night into a `tracklet'. Three tracklets over three separate nights within a 14 night window can be linked to make a discovery. Since the LSST revisits each location twice per night, these tracklets will generally be composed of pairs of observations. If the observations fulfil those requirements, the object is `discovered'. If the observations do not meet the requirements, the object is not reported. We set the linking efficiency within \texttt{Sorcha} to 0.95, which is the minimum efficiency required for the real survey.

Individual two-observations tracklets can be difficult to identify and confirm, since a single detection in the first image could correspond with a large number of detections in the second image. As such, two detections from a single night might only be linked into a tracklet days later, when another tracklet is identified and all four observations are found to satisfy a heliocentric Keplerian orbit. While this is not an issue within \texttt{Sorcha}, it will be a challenge for the real LSST, and means that two-observation tracklets might only be reported days after the observations are made. 

In about 3\% of cases, pairs will intentionally receive a third observation, potentially generating 3-observation tracklets\footnote{https://survey-strategy.lsst.io/baseline/wfd.html}. There are potential changes to the SSP which would allow it to immediately report these individual tracklets which (either by design or serendipitously) contain 3+ observations\footnote{M. Juric et al. (2022) - EU ESA Workshop on NEO Imminent Impactors Warning Coordination: https://indico.esa.int/event/422/contributions/7674/, slides 16-20, accessed Feb 4, 2026.} \citep{Wagg25}. These tracklets are easier to identify in real time, and would allow for initial orbits to be estimated. In this work, we discuss objects discovered using both the default 3-tracklet and the modified 3-observation linking methods, and refer to them as such.

\texttt{Sorcha} provides up to three outputs: (1) the full list of individual observations, (2) a summary which describes how many times each object was observed, in what filters, and whether it was linked (optional), and (3) the full ephemeris for the bodies (optional).

We use \texttt{Sorcha} version 1.0.0. See \citet{Merritt25} for more details on \texttt{Sorcha}, and \citet{Holman25} for details on the ephemeris generation.

\section{Full Survey Results}\label{sec:results}
For the `best case' survey, we assume all objects are S-type asteroids, with their higher albedo (compared to C-types) making them easier to discover. By using the same colour for all objects, we are able to directly compare objects and explore how their sizes and orbits might affect discoverability.

\subsection{Observations}\label{subsec:obs}
An `observation' is a single case of an object being within the detector footprint and being brighter than the 5-$\sigma$ limiting depth for that image. In the real LSST, these observations would be identified by difference imaging with a template image of that part of the sky and have a prompt alert issued. This single observation would contain no orbital information, but the SSP would attempt to link it to any previous detections that would satisfy a Keplerian orbit, or already-known Solar System objects \citep{Myers13, Juric17}.

Of the 216 objects injected into the \texttt{Sorcha} simulation, 30 objects are observed for a total of 130 observations. Most are observed 2 - 6 times, while four receive more than 10. The most observations of a single object is 16, obtained over a period of 20 days. Two objects are discovered by the default 3-tracklet method, and eight are discovered by the 3-observation method. The full set of observed objects is presented in Table \ref{tab:obs_full_list}.

\begin{table*}[htbp]
\centering
\caption{All objects observed by the \texttt{Sorcha} simulation. `First', `Last', and `Warning' are in days before impact. `Obs. arc' is in days. `Linked' describes whether the objects was linked by the default/three-observation linking algorithm. The `Warning' column is the time from linkage (or generation of three-observation tracklet) to impact. USG\_2024-09-04T16-39 is the imminent impactor 2024 RW$_1$. }
\label{tab:obs_full_list}

\begin{tabular}{cccccccc}
\hline
ObjID & First & Last & No. obs & Obs. arc & 3-tracklet & 3-observation & Warning time\\
\hline
USG\_2015-03-30T21-33 & 10.79 & 7.80 & 4 & 2.99 & N & N & - \\
USG\_2015-04-21T01-42 & 2.04 & 1.05 & 5 & 0.99 & N & Y & 1.84 \\
USG\_2015-04-30T10-21 & 7.34 & 4.30 & 6 & 3.04 & N & Y & 4.30 \\
USG\_2015-05-07T20-34 & 1.87 & 1.87 & 1 & 0.00 & N & N & - \\
\textbf{USG\_2015-05-10T07-4}5 & \textbf{21.18} & \textbf{1.18} & \textbf{16} & \textbf{20.00} & \textbf{Y} & \textbf{Y} & \textbf{7.22} \\
USG\_2016-03-16T23-54 & 1.91 & 1.88 & 2 & 0.02 & N & N & - \\
USG\_2016-09-14T15-01 & 4.36 & 3.36 & 3 & 1.00 & N & N & - \\
USG\_2017-02-18T19-48 & 3.60 & 1.54 & 4 & 2.06 & N & N & - \\
USG\_2017-04-30T21-28 & 3.58 & 3.56 & 5 & 0.03 & N & Y & 3.56 \\
USG\_2017-06-20T13-41 & 19.22 & 17.46 & 2 & 1.76 & N & N & - \\
USG\_2017-06-30T14-26 & 3.40 & 3.38 & 2 & 0.02 & N & N & - \\
USG\_2018-04-21T12-06 & 7.21 & 4.28 & 3 & 2.93 & N & N & - \\
\textbf{USG\_2018-09-20T18-29} & \textbf{71.49} & \textbf{1.65} & \textbf{13} & \textbf{69.84} & \textbf{N} & \textbf{N} & \textbf{-} \\
\textbf{USG\_2019-01-22T09-18} & \textbf{1262.12} & \textbf{1251.99} & \textbf{14} & \textbf{10.13} & \textbf{Y} & \textbf{Y} & \textbf{1256.02} \\
\textbf{USG\_2019-05-21T13-12} & \textbf{36.31} & \textbf{2.40} & \textbf{12} & \textbf{33.90} & \textbf{N} & \textbf{Y} & \textbf{28.29} \\
USG\_2019-09-12T02-34 & 2.86 & 2.86 & 2 & 0.00 & N & N & - \\
USG\_2020-01-16T09-31 & 4.35 & 4.35 & 1 & 0.00 & N & N & - \\
USG\_2020-01-17T21-29 & 1.86 & 1.86 & 1 & 0.00 & N & N & - \\
USG\_2020-08-30T16-08 & 2.51 & 2.49 & 2 & 0.02 & N & N & - \\
USG\_2020-10-26T15-07 & 4.49 & 4.49 & 1 & 0.00 & N & N & - \\
USG\_2021-04-13T02-16 & 4.88 & 0.92 & 3 & 3.96 & N & N & - \\
USG\_2022-03-24T03-43 & 1.82 & 0.83 & 3 & 1.00 & N & N & - \\
USG\_2022-04-12T21-59 & 1.77 & 1.74 & 2 & 0.02 & N & N & - \\
USG\_2022-07-27T04-41 & 2.97 & 2.94 & 2 & 0.02 & N & N & - \\
USG\_2023-02-14T22-55 & 5.70 & 5.68 & 4 & 0.02 & N & Y & 5.68 \\
USG\_2023-07-26T03-41 & 6.88 & 0.86 & 5 & 6.02 & N & Y & 0.86 \\
USG\_2024-07-20T14-08 & 3.46 & 3.46 & 1 & 0.00 & N & N & - \\
\textbf{USG\_2024-09-04T16-39} & \textbf{4.53} & \textbf{2.47} & \textbf{5} & \textbf{2.06} & \textbf{N} & \textbf{N} & \textbf{-} \\
USG\_2024-12-01T08-50 & 5.25 & 3.21 & 5 & 2.04 & N & N & - \\
USG\_2024-12-28T04-45 & 4.10 & 4.10 & 1 & 0.00 & N & N & - \\
\hline
\end{tabular}
\end{table*}

The mean trailed source magnitude across all observations (and filters) is 22.1. The faintest observation is 24.6 in the $r$ band, and the brightest observation is 17.7 in the $y$ band.

We take a closer look at five objects (shown in bold) - three discovered objects, one objects which are observed 10+ times but not discovered, and the 2024 RW$_1$ impactor\footnote{https://www.minorplanetcenter.net/mpec/K24/K24R68.html} \citep{Wierzchos24, Ingebretsen25}.

\subsection{Discovered Objects}\label{subsec:discovered}
\textbf{USG\_2015-05-10T07-45} was observed 16 times in four filters (\textit{rizy}) across a 20 day window (from 21 - 1 days before impact). The observations satisfied the conditions of both the default SSP (three tracklets within 14 days) and the three-observation tracklet method, so the object was `discovered' seven days before impact. 

This object came from an evolved orbit (a = 1.39 AU, e = 0.27) at low inclination (i = 2.3$^{\circ}$) and had a low impact velocity (12.2 km s$^{-1}$).

\textbf{USG\_2019-01-22T09-18} was observed 14 times in four filters (\textit{griz}) across a 10 day window, and subsequently discovered by both methods 3.5 years before impact. It had an evolved Earth-like orbit (a = 1.01 AU, e = 0.07, 2.5$^{\circ}$) and a low impact speed (11.6 km s$^{-1}$).

We note here that, because of the uncertainties associated with USG-derived orbits (Sec. \ref{subsec:usg_issues}), it is uncertain whether this specific object would have actually been observed that far before impact years pre-impact. Conversely, it is doubtful whether observations of the object would have been precise enough to predict the impact (or associate the observations with the observed event) without follow-up observations. This case is nonetheless interesting in showing that this configuration of a small metre-scale impactor detected several oppositions before impact is possible.

\textbf{USG\_2019-05-21T13-12} was observed 12 times over 34 days, and was discovered 28 days before impact using the 3-observation method. It was on a low inclination evolved orbit (a = 1.2 AU, e = 0.15, i = 2.6$^{\circ}$) and had a low impact speed (11.5 km s$^{-1}$). It has the highest impact energy (1.6 kT TNT) and absolute magnitude (H$_r$ = 28.6) of all 30 observed impactors. The fireball from this impact was also observed by members of the public in South Australia\footnote{https://www.abc.net.au/news/2019-05-22/suspected-meteor-spotted-across-the-sky-in-south-australia/11136648}.

\subsubsection{Undiscovered Objects}\label{subsec:undiscovered}
\textbf{USG\_2018-09-20T18-29} was observed 13 times over a $\sim$70 day window before impact. The observations were mostly nightly pairs, with each pair separated by 10 - 15 days, preventing the tracklets from being linked and the object from being discovered. This body was on an evolved Earth-like orbit, with a = 1.03 AU, e = 0.04 and i = 0.4$^{\circ}$. It also had a low entry velocity of around 11.1 km s$^{-1}$.

\textbf{USG\_2024-09-04T16-39} is the known imminent impactor 2024 RW$_1$ \citep{Wierzchos24, Ingebretsen25}, which was discovered approximately 11 hours before impact by the Catalina Sky Survey\footnote{https://minorplanetcenter.net/mpec/K24/K24R68.html}. Here, it was observed on five occasions 2 - 5 days before impact and not discovered
This object's orbit was not evolved (firmly in the main belt at $a$ = 2.2 AU), and it had a higher impact speed of 19.7 km s$^{-1}$.

Other events observed in this study that have been investigated previously include USG\_2017-06-30T14-26 (Baird Bay fireball over South Australia, observed by the Desert Fireball Network; \citealt{Devillepoix19}), USG\_2020-01-17T21-29 (reports from the public north of Puerto Rico, reported by the International Meteor Organisation\footnote{https://www.imo.net/asteroid-entry-north-of-puerto-rico/}), and USG\_2021-04-13T02-16 (off the coast of Florida; \citealt{Hughes22}). 

\subsection{General Trends}\label{subsec:general_trends}
From these 30 objects, we can identify general trends between the impact events, the impacting bodies, their orbits, and the \texttt{Sorcha} observations. 

\begin{figure*}
    \centering
    \includegraphics[width=1.0\linewidth]{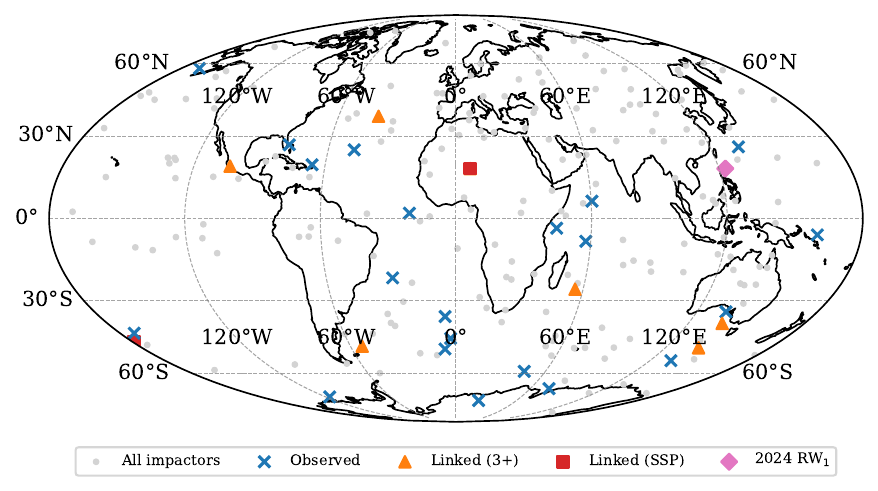}
    \caption{Map of the impact locations for the 216 USG events. Grey dots are not observed pre-impact, blue $\times$ markers are observed pre-impact, orange triangles are discovered by the 3-observation method, and red squares are discovered by the default 3-tracklet algorithm.
    The pink diamond (USG\_2024-09-04T16-39) corresponds with the known imminent impactor 2024 RW$_1$. 19 impacts are in the Souther Hemisphere, consistent with Rubin's location.}
    \label{fig:map}
\end{figure*}

\textbf{Impacts}: Of the 30 objects observed, 19 impacted into the Southern Hemisphere (compared to 11 in the Northern), and only one impacted north of 40$^{\circ}$ (Fig. \ref{fig:map}). This is consistent with Rubin's position in the Southern Hemisphere.

\begin{figure*}
    \centering
    \includegraphics[width=1.0\linewidth]{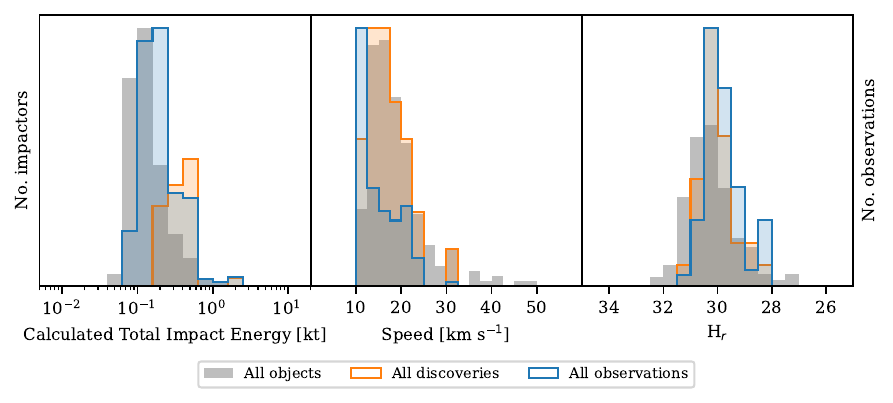}
    \caption{Comparison of all impactors (observed and not observed; grey), the total number of observations (blue), and the discovered objects (orange). (a) Impact energy: most observations (blue) are of higher-energy impactors, and all discovered objects (orange) have E $>$ $\sim$ 0.1 kT, distributed towards higher energies. (b) Speed: nearly all observations (blue) are of objects with relatively low impact speeds (v $<$ 25 km s$^{-1}$). Discovered objects (orange) generally match the input population (grey) for $V$ $<$ 25 km s$^{-1}$, but are drop off beyond that. The four bodies observed 10+ times (not shown here) all have 11.2 km s$^{-1}$ $<$ v $<$ 11.6 km s$^{-1}$. (c) H$_r$: most observations are of objects with H$_r$ $>$ 31 (D $>$ 1.5 m). The discovered objects generally have H$_r$ $>$ 31 (D $>$ 2 m). This shows that, unsurprisingly, objects that have a chance of being discovered with the USG dataset are large and slow.}
    \label{fig:impactor_three_hist}
\end{figure*}

There is a correlation between the impact speed and the number of observations obtained (Fig. \ref{fig:impactor_three_hist}). Most observations are of objects with relatively low impact speeds (v $<$ 25 km s$^{-1}$), compared to the rest of the population (which reaches 50 km s$^{-1}$). The eight discovered objects are (unsurprisingly) the larger, slower and brighter objects. Additionally, the four bodies observed 10+ times all have very low v ($<$ 11.6 km s$^{-1}$), just above the Earth's escape velocity.

The USG reports equal numbers of day vs night fireballs, while here, 26 (86.7\%) of the observed objects impacted at night. This suggests we do not observe objects which impact on the outbound section of their orbit. This is unsurprising, since ground-based telescopes can only observe objects on the night-side of the planet, but does highlight the importance of space-based observatories that can observe objects approaching the daytime side of the planet (e.g. the NEO Surveyor and NEOMIR missions; \citealt{Mainzer23, Conversi24}).

\begin{figure}
    \centering
    \includegraphics[width=1.0\linewidth]{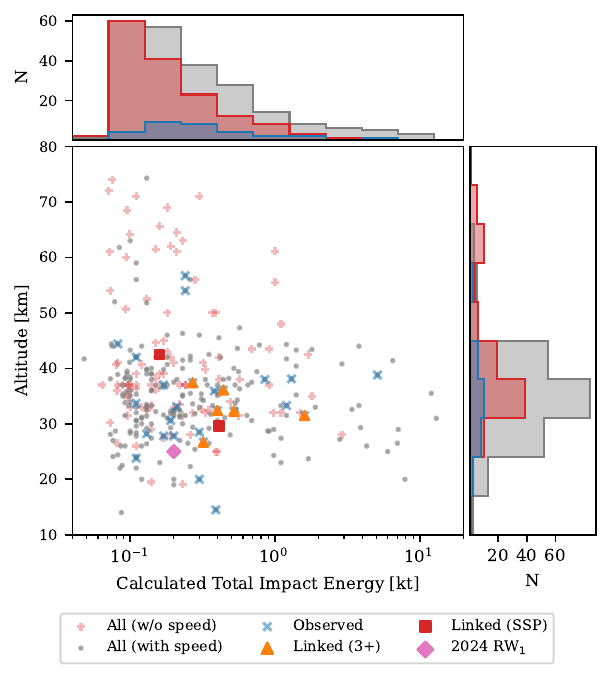}
    \caption{Calculated impact energy vs altitude for all 317 objects with reported altitudes (some of which are missing speed data), including objects without reported speeds (red $+$), objects with speeds that are not observed by LSST (grey dots) and objects with speeds that are observed (blue $\times$). Observed objects generally have low altitudes and impact energies. Speeds are rarely reported for events $>$ 50 km. (Almost) all events with E $>$ 2 kT TNT have reported speeds.} 
    \label{fig:kt_alt}
\end{figure}

None of the observed impactors generated fireball events with peak brightness altitudes $>$ 60 km, and only one had impact energy $>$ 2 kT TNT (Fig. \ref{fig:kt_alt}). We also note that the USG data rarely report speeds for events above 50 km, and almost always reports speeds for events with impact energies $>$ 2 kT TNT. 

\textbf{Impactors}: USG sensors are sensitive to events with impact energies $>$ 0.05 kT, corresponding to 1 m objects with H$_r$ $\sim$ 32 (assuming albedo of 0.24). We find that LSST primarily observes impactors with H$_r$ $>$ 31 (D $>$ 2 m; Fig. \ref{fig:impactor_three_hist}). The impactors observed 10+ times all have H$_r$ $>$ 30.5 (D $>$ 2.5 m). 

\begin{figure*}
    \centering
    \includegraphics[width=1.0\linewidth]{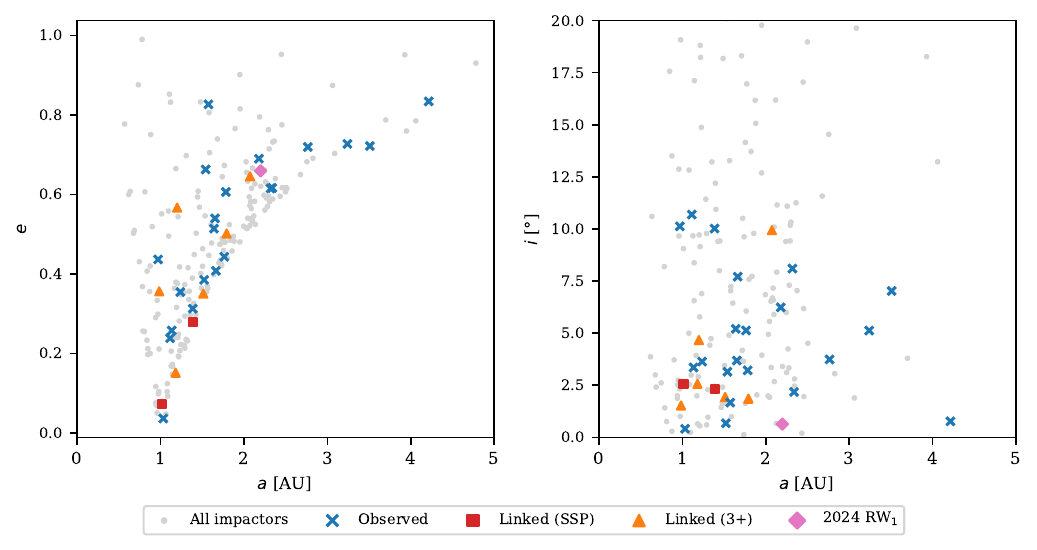}
    \caption{Semimajor axis compared to eccentricity and inclination. Grey dots are not observed pre-impact, blue $\times$ markers are observed pre-impact, orange triangles are discovered by the 3-observation method, and red squares are discovered pre-impact by the default 3-tracklet method. The pink diamond (USG\_2024-09-04T16-39) corresponds with the known imminent impactor 2024 RW$_1$. The 3-tracklet discovered objects are on more evolved orbits than the 3-observation discovered objects, suggesting the modified linking algorithm can access a larger orbital range of objects.}
    \label{fig:sma_e_i}
\end{figure*}

\textbf{Orbits}: Fig. \ref{fig:sma_e_i} shows the orbital elements derived from the USG velocity data (speed and direction) of the impacts. There is a notable cut-off of observed objects around $i$ = 10$^{\circ}$, above which none are observed. As noted in the previous subsection, the two objects discovered by the default 3-tracklet method have low $a$ ($<$ 1.5 AU), $e$ ($<$ 0.3) and $i$ ($<$ 2.5$^{\circ}$). The two objects linked by the 3-observation method are on slightly less evolved orbits, which is consistent with that method only requiring observations across one night (and thus not requiring the object to be on an Earth-like orbit). 

\begin{figure}
    \centering
    \includegraphics[width=1.0\linewidth]{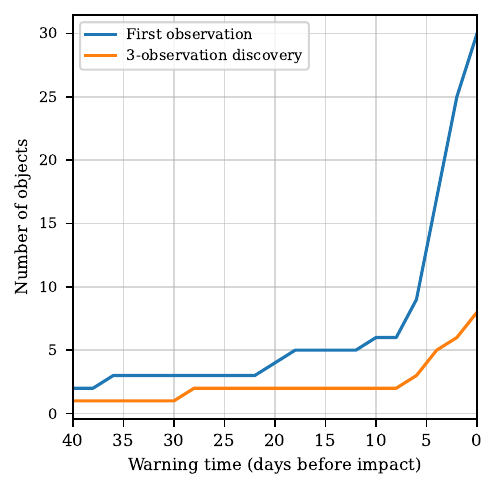}
    \caption{Days before impact of the first observations for the observed bodies (blue), and of objects discovered by the default SSP or 3-observation tracklet method (orange). Most objects are first observed with a week of impact. One object (USG\_2019-01-22T18-29), discovered over three years in advance, is cut off. The median time for the 30 objects' first observation is 4.3 days pre-impact, and eight discoveries are made 5.0 days pre-impact. For reference, the earliest current warning time is 21 hours (2014 AA; \citealt{Kowalski14, Farnocchia16})}.
    \label{fig:lead_times}
\end{figure}

\textbf{Observations}: For all observed bodies, the median time before impact of the \textit{first} observation is 4.3 days (Fig. \ref{fig:lead_times}). Most objects with 2 observations have observation arcs of $\sim$0.5 hours, which corresponds to LSST's half hour revist time. 15 objects have observation arcs $>$ 1 day. Eight objects are discovered between 0.8 - 28 days pre-impact, for a median warning time of 5.0 days (excluding USG\_2019-01-22T09-18, discovered 3.5 years before impact).

\section{Discussion}\label{sec:discussion}
\subsection{Results Summary}\label{subsec:results_summary}
We send \texttt{Sorcha} 216 metre-scale USG objects across a 10 year span. 30 objects (13.9\%) are observed pre-impact, with most receiving 2 - 4 observations. Four objects (1.9\%) receive 10+ observations (three of which are discovered). Two objects (0.9\%) are discovered with the default 3-tracklet linking algorithm and eight (3.7\%) are linked by the 3-observation method (see Section. \ref{subsec:SSP}).

\subsection{Linking, Discoveries, and the SSP}\label{subsec:linking_discoveries}
We find that the LSST will be able to make observations of metre-sized impactors pre-impact. However, most of these observations will not be linked by the SSP and/or discovered. This is primarily due to the survey strategy that the LSST employs, compared to dedicated NEO and planetary defence surveys. 

Most dedicated NEO surveys image sections of the sky $\sim$4 times each night, with observations being separated by 30 - 60 minutes \citep{Tonry18, Chambers16, Drake09}. Detections within these images can be immediately linked to generate four-observation tracklets, allowing for reasonable estimations of the object's orbit and for the tracklet to be compared to known Solar System objects. If the tracklet fails to match any known objects, it is sent to the NEO Confirmation Page (NEOCP\footnote{https://minorplanetcenter.net/iau/NEO/toconfirm\_tabular.html}). There, programs like NASA/JPL's SCOUT and ESA's Meerkat use systematic ranging techniques \citep{Drury26, Chesley04, Farnocchia15, Keys19} to flag objects with high impact probabilities, which are prioritised for follow up observations.

In contrast, LSST makes two visits per night, and so (mostly) generates two-observation tracklets. These tracklets will not be reported to the NEOCP, since they are a) difficult to identify across a single night and b) would risk flooding the NEOCP with unconstrained orbits (\citealt{Wagg25}; see Section \ref{subsec:SSP}). All discoveries will require the linking of three tracklets which, based on LSST's three-day cycling survey strategy, will take on the order of a week. This delay will prevent many imminent impactors from being discovered (our median time of earliest observation was 3.5 days before impact). 

Proposed changes to the SSP would allow it to report tracklets with 3+ candidates\footnote{M. Juric et al. (2022) - EU ESA Workshop on NEO Imminent Impactors Warning Coordination: https://indico.esa.int/event/422/contributions/7674/, slides 16-20, accessed Feb 4, 2026.} \citep{Wagg25}, either in real time or during the following day, in a similar way to current impact-hunters. In this work, we find four objects which had single tracklets with 3+ observations, which would have been discovered if these modifications to the SSP were applied. These tracklets were collected between 20 hours and 28 days pre-impact (excluding USG\_2019-01-22T09-18, 3.5 years pre-impact). Implementing these changes to the SSP could significantly increase the number of metre-scale imminent impactors discovered. 

\subsection{Lead Times}\label{subsec:lead_times}
Our eight discoveries are made between 0.8 - 28 days pre-impact, with a median value of 5.0 days (excluding the object linked over three years before impact, which would have been extremely difficult to identify). Most objects (discovered and undiscovered) are first observed several days (median = 4.3) before impact, with some extending to weeks before impact.

For comparison, the longest lead time of the 11 known imminent impactors is 21 hours (2014 AA; \citealt{Kowalski14, Farnocchia16}), and the mean is $\sim$9 hours. Discovering impactors days to weeks in advance - even if rare - will provide incredible potential for dedicated follow-up surveys of the object, both with telescopes and in the potential deployment of ground based sensors to observe the fireball (e.g. \citealt{Clemente25}). These warning times also begin to align with the requirements for planetary defence responses, so would make for useful test cases \citep{Brown13, Koschny24}.

\begin{figure}
    \centering
    \includegraphics[width=1.0\linewidth]{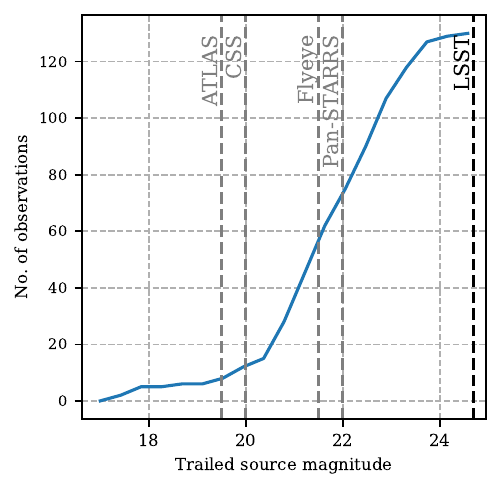}
    \caption{Trailed source magnitude of observations from this work, compared to the limiting magnitudes of ATLAS ($c$-band; \citealt{Tonry18}), CSS (V-mag; \citealt{Drake09}), Pan-STARRS (V-band; \citealt{Chambers16, Wainscoat21}), Flyeye (V-mag; \citealt{Cibin19, Fohring24} and LSST ($r$ band; \citealt{Ivezic19}). ATLAS and CSS would not have made most of the observations that we simulate here, while Flyeye and Pan-STARRS would have made only half. The other $\sim$half (fainter than magnitude 22) are only accessible to LSST. This does not account for cadence differences between the surveys, which is a driving factor in whether observations are linked and discoveries are made.}
    \label{fig:app_mag}
\end{figure}

Most of this increase in warning time is due to the LSST's deeper single-exposure limiting magnitude compared to other NEO-hunting surveys. Fig. \ref{fig:app_mag} shows the magnitude of all observations made in this study compared to the limiting magnitudes of ATLAS ($c$-band $\sim$ 19.5; \citealt{Tonry18}), CSS (V-mag $\sim$ 20; \citealt{Drake09}), Pan-STARRS (V-band $\sim$ 22; \citealt{Chambers16, Wainscoat21}), Flyeye (V-mag $\sim$ 21.5; \citealt{Cibin19, Fohring24}) and LSST ($r$-band $\sim$ 24.7; \citealt{Ivezic19}). ATLAS and CSS could have only made the observations brighter than magnitude 20 (and so missed most of the observations here), while Flyeye and Pan-STARRS would have missed about half of the observations. LSST, with its deeper limiting magnitude, will be able to see smaller objects further from the Earth (and thus earlier before impact) than pre-existing surveys. While the LSST cadence is not optimal for discovering impactors, those which is does discover will be significantly earlier. As such, the LSST is expected to work in tandem with other surveys, rather than replace them.

\subsection{Precoveries}\label{subsec:precoveries}
We make 63 observations of 22 unique objects which are \textit{not} discovered using either the current 3-tracklet or modified 3-observation linking methods. These observations are made days (or weeks) in advance, and will still be valuable, even if the impactors are not discovered by the LSST.

If these objects are then detected by dedicated impactor-hunting surveys and reported to the Minor Planet Centre (MPC), the SSP will associate past LSST observations with the new object for as far back as reasonable for the accuracy of the orbit \citep{Bellem20}. These precoveries could extend the observational arc dramatically and assist with making follow-up observations and constraining the impact location. 

In other cases, LSST may make observations of bodies which remain undiscovered until they impact the Earth. If the impact itself is observed with enough accuracy by either the USG or other ground-based networks, a pre-impact ephemeris for the object could be generated and searched for occasional, unreported observations by LSST (potentially even below the 5-$\sigma$ level).
\citet{Clark23} have shown that telescope data precovery from fireball-derived ephemeris is possible. 

\section{Future LSST Predictions}
\label{sec:predictions}

\subsection{Secondary Survey Results}
\label{subsec:split_survey}
In the previous section, we assumed every object was an S-type in order to give it the best change of being discovered. To perform a more realistic survey in order to make predictions, we run a second simulation assuming all objects are C-types (a `worst case' scenario). We then iterate 500 surveys by randomly assigning an even split of S- and C-types for each survey.

We take the mean values for the number of observations, number of observed objects, and number of discovered objects, and the median for the warning time (to minimise the effects of USG\_2019-01-22T09-18, detected 3.5 years before impact when simulated as an S-type). We take our uncertainties as the standard deviation, acknowledging that the real values could be higher or lower depending on the true S-/C-type split.

Assuming the even distribution of S- and C-type objects, we find 19.5 $\pm$ 1.2 impactors observed for a total of 72 $\pm$ 10 observations. 1.0 $\pm$ 0.7 objects are linked by the default 3-tracklet SSP, and 4.5$\pm$1.1 by the 3-observation method, for a median warning time of 4.3 $\pm$ 3.2 days.

\subsection{Scaled Total Predictions}
\label{subsec:total_predictions}
We assume the USG currently detects all impacts of metre-scale bodies \citep{Brown02, Devillepoix19}. To derive the orbit for an object, and simulate it in this work, we require its impact speed and radiant. However, of all 372 events recorded between 1 January 2015 - 31 December 2024, only 216 (58.1\%) have reported speeds. In this work, we have used these 216 events.

To make predictions for the full metre-scale impacting population, we scale our reported values (and uncertainties) by a factor of $\frac{372}{216} = 1.72$.

Based on this, we predict LSST to observe 34 $\pm$ 2 (up from 19.5) objects pre-impact over the 10 year period, and make 125 $\pm$ 18 total observations (up from 72). We predict 2 $\pm$ 1 discoveries of imminent impactors with the default SSP (up from 1), and 8 $\pm$ 2 (up from 4.5) discoveries if the modified SSP is introduced (and assuming every 3-observation tracklet is reported and followed up by other observatories). The lead time remains the same, at 4.3 $\pm$ 3.2 days.

\section{Conclusions}\label{sec:conclusions}
In this work, we use the LSST Solar System survey simulator \texttt{Sorcha} \citep{Holman25, Merritt25} to investigate whether the LSST would have observed and/or discovered any of the metre-scale meteoroids that impacted the Earth over the last decade. We use data obtained by the USG sensors and available on the CNEOS\footnote{https://cneos.jpl.nasa.gov/fireballs/} website over the period 2015 January 1 - 2024 December 31. 

We performed two simulations, one `best case' assuming all objects were bright S-types, and one assuming a more realistic 50/50 distribution between S-and C-types. We inspect the individual objects from the first simulation, and make predictions based on the second.

Our main conclusions are as follows:

\begin{itemize}
    \item Of the 216 objects from the `best case' simulation, the LSST would have observed 30 impactors for a total of 130 observations. 
    \item Of these 30 objects, two were linked by the default 3-tracklet linking algorithm. Eight were discovered by a custom method which uses single 3-observation tracklets, and which is a proposed change to the current Solar System Processing pipeline. 
    \item These eight discoveries were made 5 days pre-impact (median), significantly earlier than the current longest lead time for a real imminent impactor of 21 hours for 2014 AA \citep{Kowalski14, Farnocchia16}.
    \item Based off the `realistic' simulation, and accounting for limitations in the USG CNEOS dataset, we predict the LSST will observe 34 $\pm$ 2 unique metre-scale impactors for 125 $\pm$ 18 individual observations, and discover 8 $\pm$ 2 pre-impact over the course of its 10 year survey (assuming the 3-observation method is applied). We predict a median warning time of 4.3 $\pm$ 3.2 days. These values are consistent with the lower end of previous estimates (e.g. \citealt{Juric23}) and other recent work (\citealt{Cheng26} and references therein). 
    \item Even with the SSP changes, we still predict 26 $\pm$ 2 objects will be observed a few (e.g. 1 - 4) times but remain undiscovered, based on limitations of the linking algorithms. However, if these objects are then discovered by other surveys (either telescopic or fireball-based), these observations could be linked \citep{Bellem20, Clark23}, which would significantly extend observation arcs and improve orbit determination and impact location predictions.
\end{itemize}

While this paper was under review, a preprint by \citet{Chow26} was released examining the detectability of Earth impactors by the LSST using \texttt{Sorcha}, assuming a modified linking algorithm using two streaked detections in the same night. 
Their discoveries (and predictions) are slightly higher than ours, with the different linking techniques capturing different objects. This supports the notion that current predictions (both theirs and ours) are likely lower limits, and that multiple approaches will be required to make the most of LSST data.

The LSST providing a sample of a few tens of objects observed pre-impact would allow for the comparison of impact-derived masses with pre-impact photometry. This could begin to uncover systematic issues with average densities and albedos that we had to assume in this work, and in turn allow for better predicts of the size of impactors based on pre-impact observations.

While Rubin and the LSST will likely not become the leading facility for pre-impact discoveries, the $\sim$1 discovery per year that we do predict will come with several days warning, rather than the current average of a few hours. This will allow ample time for follow-up observations, and potentially the deployment of specialised observation equipment for the fireball phase, which is key for planetary defence contexts and the continued study of the meteoroid-fireball-meteorite connection.
 
\begin{acknowledgments}
We would like to thank Peter Yoachim for providing the custom retrospective \texttt{Sorcha} pointing database, and Ian Chow and Mario Juric for extremely useful discussions. We also thank the anonymous reviewer for their useful comments. We also thank the anonymous reviewer for their useful comments.

MF and SD are were supported by the Commonwealth through an Australian Government Research Training Program Scholarship.

This research has made use of data and/or services provided by the International Astronomical Union's Minor Planet Center. 
This research has made use of NASA’s Astrophysics Data System Bibliographic Services. 

Some of the results in this paper have been derived using the healpy and HEALPix\footnote{HEALPix (http://healpix.sf.net)} packages.

This material or work is supported in part by the National Science  Foundation through Cooperative Agreement AST-1258333 and Cooperative Support Agreement AST1836783 managed by the Association of Universities for Research in Astronomy (AURA), and the Department of Energy under Contract No. DE-AC02-76SF00515 with the SLAC National Accelerator Laboratory managed by Stanford University.

AI tools were used to aid in coding, but did not contribute to the scientific content, analysis or conclusions of this work. The authors take all responsibility for the work presented here.

\end{acknowledgments}

\begin{contribution}
MF conducted the research and wrote the manuscript.
HD assisted with editing, writing and discussions, and provided supervision.
SD assisted with editing, writing and discussions.
\end{contribution}


\software{\texttt{Sorcha} \citep{Merritt25, Holman25}, ASSIST \citep{2023PSJ.....4...69H,hanno_rein_2023_7778017}, Astropy\footnote{http://www.astropy.org} \citep{2013A&A...558A..33A,2018AJ....156..123A,2022ApJ...935..167A}, Healpy \citep{Zonca2019,2005ApJ...622..759G}, Matplotlib \citep{Hunter:2007}, Numpy \citep{harris2020array}, pandas \citep{mckinney-proc-scipy-2010, pandas:2025}, REBOUND \citep{Rein12,Rein15}, sqlite (\url{https://www.sqlite.org/index.html}), sqlite3 (\url{https://docs.python.org/3/library/sqlite3.html}), Jupyter Notebooks \citep{soton403913}.}

\bibliography{Sorcha_citations,sample701,DFN_library}{}

\bibliographystyle{aasjournalv7}

\end{document}